\documentclass{article}
\usepackage{spconf,amsmath,graphicx,amssymb,booktabs}
\usepackage{color}

\title{A two-step backward compatible fullband speech enhancement system}
\name{Xu Zhang, Lianwu Chen, Xiguang Zheng, Xinlei Ren, Chen Zhang, Liang Guo, Bing Yu }
\address{Kuaishou Technology, Beijing, China}
\begin{document}
%
\maketitle
\vspace{-9pt}
\begin{abstract}

Speech enhancement methods based on deep learning have surpassed traditional methods. While many of these new approaches are operating on the wideband (16kHz) sample rate, a new fullband (48kHz) speech enhancement system is proposed in this paper. Compared to the existing fullband systems that utilize perceptually motivated features to train the fullband speech enhancement with a single network structure, the proposed system is a two-step system ensuring good fullband speech enhancement quality while backward compatible to the existing wideband systems. 
\end{abstract}
\vspace{-9pt}
\section{Introduction}
\vspace{-3pt}
\label{sec:intro}

In recent years, deep learning based speech enhancement (SE) approaches have achieved significant improvement over the traditional signal processing based methods and become the main stream. While recent systems \cite{8547084,ValinIPGHK20,westhausen_dual-signal_2020,li_icassp_2021,li21g_interspeech, zhang21t_interspeech} are performed well for the wideband (16kHz) speech \cite{DNS20211,DNS20212}, only few of them \cite{8547084,ValinIPGHK20} have considered the fullband (48kHz) speech processing. Extending the speech bandwidth from wideband to super-wideband (32kHz) and fullband (48kHz) has widely practiced in contemporary speech codecs \cite{opus,7179063} demonstrating improved speech quality \cite{5273816}. The fullband speech compression system is usually built on-top of the existing wideband speech compression methods to ensure scalability under limited bitrates and backward compatibility of the existing wideband speech systems. In the remainder of this paper, the sampling frequency is abbreviated to '16k', '32k' and '48k' to avoid confusion with the signal frequency.

For the fullband SE systems, employing the psychoacoustically motivated features (such as the Bark-frequency cepstral coefficients (BFCC) in \cite{8547084} and equivalent rectangular bandwidth (ERB) features in \cite{ValinIPGHK20}) can significantly reduce the frequency-wise feature dimension and thus reduce the overall complexity of the fullband system. The drawback of using these features is the smeared frequency resolution in the high frequency region (8-24kHz). This leads to degraded performance in the high frequency region especially for the SE problem where the signal to noise ratio (SNR) in the high frequency region is usually much lower than the low frequency region (0-8kHz) as discussed in this paper. 
\begin{figure}
\vspace{-9pt}
\begin{minipage}[b]{1\linewidth}
  \centering
  \centerline{\includegraphics[width=8cm]{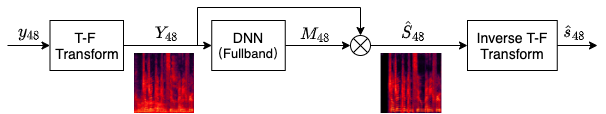}}
  \centerline{(a) the existing one-step structure}\medskip
\end{minipage}
\hfill
\begin{minipage}[b]{1\linewidth}
  \centering
  \vspace{-3pt}
  \centerline{\includegraphics[width=8.5cm]{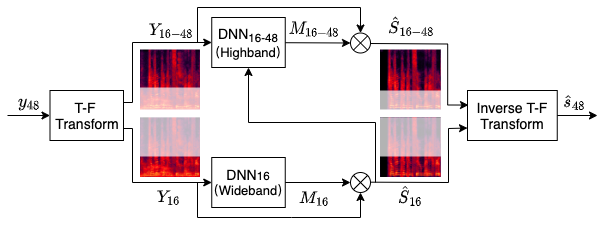}}
  \centerline{(b) the proposed two-step structure}\medskip
\vspace{-15pt} 
\end{minipage}
\caption{Fullband speech enhancement architectures.}
\label{fig:NetworkArchitecture}
\vspace{-15pt} 
\end{figure}

To solve these problems, a two-step backward compatible fullband speech enhancement system is proposed. Compared to \cite{8547084,ValinIPGHK20}, the contribution of the proposed method can be summarized by: a) the two-step system is proposed consisting of a pre-trained wideband (0-8kHz) SE system (the first step) and a highband (8-24kHz) SE system (the second step). Backward compatibility is ensured as any existing wideband SE systems can be used. b). the highband SE system takes the estimated wideband clean speech as the additional input alongside with the 8-24kHz noisy speech to provide extra guidance.  Another practical benefit for this arrangement is that the publicly available speech and noise datasets are commonly released under 16k sample rate. Hence, dividing the fullband SE system into a wideband and a highband subsystem can optimally utilize the available resources ensuring good generalization ability. 

The proposed system is evaluated using two commonly used SE structures and compared with different one-step fullband training strategies using both psychoacoustically motivated and the short-time Fourier transform (STFT) features. As shown in the objective and subjective results, the proposed two-step system outperforms the other conditions. 


\vspace{-6pt}
\section{RELATED WORKS}
\vspace{-3pt}
\label{sec:system description}

\subsection{Problem formulation}
\label{ssec:Problem Formulation}
\vspace{-3pt}
A time domain noisy-reverberant mixture $y(t)$ is formed by: 
\begin{equation}
\label{eq:signal model}
y(t)=h(t)*s(t)+n(t) 
\end{equation}
where $s(t)$ is the speech signal, $h(t)$ is the transfer function from the talker to the microphone, $*$ denotes the convolution, and $n(t)$ is the noise signal. The purpose of the SE system is to estimate $s(t)$ from $y(t)$ by removing $n(t)$ and $h(t)$.

\vspace{-6pt}
\subsection{One-step fullband SE system}
\vspace{-3pt}
\label{ssec:one-step-sys}
Figure \ref{fig:NetworkArchitecture} (a) presents the existing one-step SE system. The time-frequency (T-F) transform is performed on the input noisy speech $y$ to obtain the T-F domain representation $Y$. A common choice of the T-F domain features for a wideband SE system is the STFT feature as in \cite{westhausen_dual-signal_2020,li_icassp_2021,li21g_interspeech, zhang21t_interspeech}, and psychoacoustically motivated features such as the BFCCs in \cite{8547084} and the ERB features in \cite{ValinIPGHK20}. For the existing one-step fullband SE systems, the T-F domain noisy-reverberant speech $Y_{48}$ is directly fed into a deep neural network (DNN) to estimate an ideal ratio mask or an ideal complex mask $M_{48}$. The T-F domain estimated clean speech $\hat{S}_{48}$ can be obtained by,
\begin{equation}
\vspace{-3pt}
  \hat{S}_{48} = Y_{48} \cdot M_{48}
  \label{eq:one_step_est}
\end{equation}
The time-domain estimated clean speech $\hat{s}_{48}$ is formed using the inverse T-F transform, such as using the Inverse short-time Fourier transform.

\vspace{-6pt}
\section{PROPOSED APPROACH}
\vspace{-3pt}
\label{sec:PROPOSED APPROACH}

\subsection{Two-step fullband SE system}
\begin{figure}
  \centering
  \includegraphics[width=0.8\linewidth,height=3.2cm]{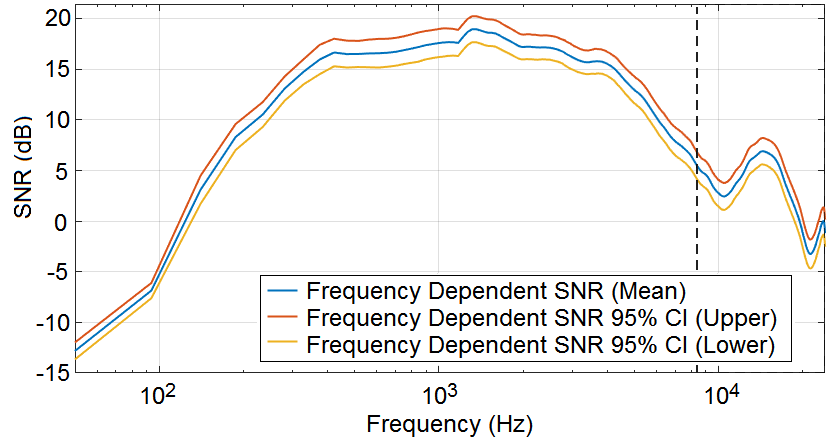}\\
  \vspace{-8pt}
  \caption{Frequency dependent SNR for the VCTK 48k testset}
  \label{fig:figVCTK}
\end{figure}

\vspace{-3pt}
\label{ssec:two-step-sys}
The proposed two-step fullband SE system is shown in Figure \ref{fig:NetworkArchitecture} (b). After the T-F transform, $Y_{48}$ is divided into two parts, namely, $Y_{16}$ and $Y_{16-48}$, representing the wideband and highband time-frequency component below 8kHz and between 8kHz and 24kHz, respectively. The T-F domain estimated clean speech $\hat{S}_{48}$ can be obtained by,
\begin{align}
\vspace{-2pt}
  \hat{S}_{48} &= concat(\hat{S}_{16},\hat{S}_{16-48}) \\
  &= concat(Y_{16} \cdot M_{16}, Y_{16-48} \cdot M_{16-48}) 
  \vspace{-3pt}
  \label{eq:two_step_est1}
\end{align}
where $\hat{S}_{16}$ is the final output signal of the standalone wideband SE  network (DNN$_{16}$), $M_{16}$ is the intermediate mask. Backward compatibility is ensured
as any existing wideband SE networks can be employed, such as SE networks with the encoder-decoder structure \cite{li_icassp_2021,li21g_interspeech, zhang21t_interspeech} and the stacked LSTM structure \cite{8547084,ValinIPGHK20,lstm}. Compared  to  the  existing  one-step systems which require 48k datasets, datesets with 16k sample rate can be fully utilized to ensure good generalization of the wideband SE. $M_{16-48}$ is the mask of the higher frequency bands $Y_{16-48}$, which can be obtained by,
\begin{equation}
\vspace{-2pt}
  M_{16-48} = DNN_{16-48}(Y_{16-48}, \hat{S}_{16})
  \label{eq:one_step_est}
\end{equation}
The estimated wideband speech signal $\hat{S}_{16}$ is fed to the highband SE network (DNN$_{16-48}$) alongside with $Y_{16-48}$. Using the estimated signal $\hat{S}_{16}$ to assist producing $\hat{S}_{16-48}$ can significantly improve the highband SE quality.  
Figure \ref{fig:figVCTK} shows the frequency dependent SNR (fSNR) of the fullband noisy speech signals based on the VCTK \cite{VCTK} test set, where the red and the yellow curves indicates 95\% confidence interval (CI). As shown in the right side of the dashed vertical line, the highband (8-24kHz) region has significantly lower fSNRs compared to the wideband (0-8kHz) region (left side of the vertical line). This demonstrates the motivation of employing $\hat{S}_{16}$ to assist the $\hat{S}_{16-48}$ estimation as shown in Figure \ref{fig:NetworkArchitecture} (b).


\vspace{-9pt}
\subsection{Highband SE network}
\label{ssec:proposedDNN1648}
\vspace{-6pt}
\begin{figure}[t]
  \centering
  \includegraphics[width=1.1\linewidth]{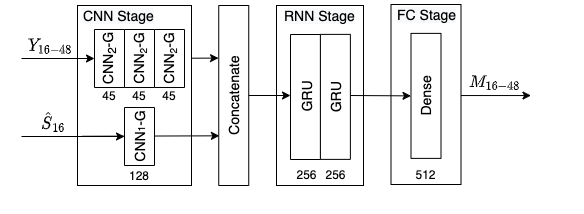}\\
  \vspace{-15pt}
  \caption{The proposed CRNN-based highband SE network}
  \label{fig:DNN1648}
  \vspace{-15pt}
\end{figure}

As shown in Figure \ref{fig:NetworkArchitecture} (b) and \ref{fig:DNN1648}, the proposed highband SE network DNN$_{16-48}$ is based on the convolution recurrent network (CRN) \cite{tan2018convolutional} that serialized the Convolutional Neural Network (CNN) and the Recurrent Neural Network (RNN). 

The highband frequency component $Y_{16-48}$ is the input to the upper CNN stream containing three 2-dimensional CNN groups (CNN$_{2}$-G) to extract the highband features. Each CNN$_{2}$-G is consisted by a 2D convolutional layer producing the feature maps with the 2D convolutional filters. The feature maps are fed to the exponential linear unit (ELU) activation function followed by a batch normalization layer and a dropout layer with the dropout rate set to 0.25. The input to the lower CNN stream of the CNN Stage is the estimated wideband speech $\hat{S}_{16}$ obtained from the pre-trained DNN$_{16}$ network. The lower CNN stream contains a 1-dimensional CNN group formed by the identical structure as in CNN$_{2}$-G except the 2D-CNN is replaced by a 1D-CNN. The output of the upper 2D-CNN stream is reshaped and concatenated with the output of the lower CNN stream, and then fed into the recurrent layers. A dropout layer with 0.25 dropout rate is also employed after each recurrent layer. The final output tensor is produced using the output of the last recurrent layer by a feed-forward layer with the sigmoid activation function. Details of the hyper-parameters are presented in Table \ref{tab:networkp}.


\begin{table}
  \caption{Hyper-parameters for proposed DNN$_{16-48}$}
  \label{tab:networkp}
  \centering
  \small
  \begin{tabular}{l|ccc c c}
    \toprule
    {\textbf{Layer}} &
    \multicolumn{3}{c}{\textbf{CNN}} & 
    \multicolumn{1}{c}{\textbf{RNN}} & 
    \multicolumn{1}{c}{\textbf{FC}} \\ 
     & \textbf{Channels} & \textbf{Size} & \textbf{Stride} & \textbf{Units}  & \textbf{Units} \\
     \hline
     Conv2D     & 45 & [4,3] & [1,2] & &  \\
     \hline
     Conv2D     & 45 & [1,3] & [1,2] & &  \\
     \hline
     Conv2D     & 45 & [1,3] & [1,2] & &  \\
     \hline
     Conv1D     & 128 & 1 & 1 & &  \\
     \hline     
     GRU     &  &  &  & 256 & \\
     \hline
     GRU     &  &  &  & 256 & \\
     \hline
     Dense     &   &   &   &  &  512 \\
     \bottomrule
  \end{tabular}
  \vspace{-10pt}
\end{table}
\vspace{-10pt}

\subsection{Loss function}
\vspace{-6pt}
\label{ssec:Neural Network}
As proposed in our earlier work \cite{zhang21t_interspeech}, the ideal amplitude mask (IAM) weighted Mean Absolute Logarithmic Error (MALE) loss is applied to better suppress the noise for the low-SNR time-frequency bins. The loss is  calculated by,  
\vspace{-10pt}
\begin{equation}
\begin{split}
\label{eq:IAM_loss}
\mathcal{L}_\text{IAM-MALE} = \sum_{t}\sum_{f} W_\text{IAM}(t,f) \cdot  \qquad \quad\quad\quad\quad \\
|ln(X'_{mag}(t,f)+1) -ln( X_{mag}(t,f)+1)|
\vspace{-15pt}
\end{split}
\end{equation}
where,
\begin{equation}
\label{eq:IAM_Weight}
W_\text{IAM}(t,f) = e^{a/(b+M(t,f))}
\vspace{-3pt}
\end{equation}
\begin{equation}
\label{eq:IAM_defind}
M(t,f)=(\frac{X_{mag}(t,f)}{Y_{mag}(t,f)})^\gamma
\vspace{-3pt}
\end{equation}
$X'_{mag}(t,f)$, $X_{mag}(t,f)$ and $Y_{mag}(t,f)$ are the predicted, clean and noisy speech amplitude spectrum, respectively. $M(t,f)$ is the IAM. According to \cite{zhang21t_interspeech}, the value of $\gamma$, $a$ and $b$ are set to 1, 2 and 1, respectively.

\vspace{-6pt}
\section{Datasets, Experiments and Results}
\label{sec:datasets, experiments and results}
\vspace{-6pt}
\label{ssec:datasets}
\subsection{Datasets}
\vspace{-6pt}

The 48k clean speech and noise dataset of the experiments is from the DNS Challenge (INTERSPEECH 2021) which contains multiple languages and various transient and stationary noise types. We synthesize the training data with randomly selected SNRs chosen from -5 to 30 dB. To further improve the robustness in reverberated environments, the noisy and target signals are convolved with measured or simulated room impulse responses. Various EQ filters are also applied to simulate the frequency response of different microphones. To avoid speech distortion introduced by dereverberation, speech with 75ms early reflection is used as training target. This results 400 and 100 hours of noisy speech signals for training and validation. The 48k VCTK \cite{VCTK} test set containing 824 sentences under different SNRs is employed for testing. 

\begin{table}
  \caption{The Encoder-Decoder Structure}
  \label{tab:encdec}
  \centering
  \small
  \begin{tabular}{l|ccc c c}
    \toprule
    {\textbf{Layer}} &
    \multicolumn{3}{c}{\textbf{Encoder (Decoder)}} & 
    \multicolumn{1}{c}{\textbf{RNN}} & 
    \multicolumn{1}{c}{\textbf{FC}} \\ 
     & \textbf{Channels} & \textbf{Size} & \textbf{Stride} & \textbf{Units}  & \textbf{Units} \\
     \hline
     Conv2D     & 45 & [4,3] & [1,2] & &  \\
     \hline
     Conv2D     & 45 & [1,3] & [1,2] & &  \\
     \hline
     Conv2D     & 45 & [1,3] & [1,2] & &  \\
     \hline
     GRU     &  &  &  & 256 & \\
     \hline
     GRU     &  &  &  & 256 & \\
     \hline
     de-Conv2D    & 8 & [1,5] & [1,2] &  & \\
     \hline
     de-Conv2D   & 1 & [1,3] & [1,1] &  &  \\
     \hline
     Dense     &   &   &   &  &  D \\
     \bottomrule
  \end{tabular}
  \vspace{-6pt}
\end{table}
  \vspace{-10pt}

\subsection{Experimental setups}
\label{ssec:expsetup}

Four one-step baseline systems are compared in the experiments. Condition FFT$_{768}$ is a one-step system (Figure \ref{fig:NetworkArchitecture} (a)) that employs the T-F representation of the input 48k noisy speech signal using a 1536-point STFT with 480 point stride (769 frequency bins for each frame). Conditions Mel$_{48}$, Mel$_{64}$, Mel$_{80}$ are all one-step systems with the settings identical to Condition FFT$_{768}$, except the mel-scaled spectrogram with corresponding resolutions (i.e. 48, 64 and 80). It should be noted that condition Mel$_{48}$ is designed to provide similar psycho-acoustically motivated frequency resolution compared to the existing works in \cite{8547084,ValinIPGHK20}. Condition Mel$_{80}$ is selected as it is a common choice for fullband Text to Speech (TTS) \cite{ping2018deep} and singing voice synthesis \cite{Chen2020HiFiSingerTH}. 

Three two-step systems are evaluated. 1536-point STFT with 480-point stride is employed. Them resulting 769 frequency bins are divided into a wideband part (1-257 frequency bins) and a highband part (258-769 frequency bins). Condition TS-FFT$_{768}^{e16k}$ is implemented using the two-step structure as in Figure \ref{fig:NetworkArchitecture} (b) with the aid of estimated 16k wideband signal $\hat{S}_{16}$ for highband SE. Condition TS-FFT$_{768}$ is the condition TS-FFT$_{768}^{e16k}$ without the aid of $\hat{S}_{16}$, and condition TS-FFT$_{768}^{n16k}$ is the the condition TS-FFT$_{768}^{e16k}$ with the estimated $\hat{S}_{16}$ replaced by noisy $Y_{16}$.

We use the Adam optimizer to minimize the IAM weighted loss for all systems. The learning rate is set to 0.001 and decays by multiplying a factor of 0.5 when the validation loss does not decrease for 5 epochs. Two kinds of SE structures are studied in the experiments for the one-step fullband DNN systems and the two-step wideband DNN$_{16}$ systems. 
\begin{itemize}
  \vspace{-3pt}
    \item Exp1: Both fullband DNN and wideband DNN$_{16}$ employ the encoder-decoder structure proposed in \cite{zhang21t_interspeech}. The hyper-parameters are shown in Table \ref{tab:encdec}, where the number of FC units (D) depends on the corresponding frequency dimension for each condition.
    \vspace{-6pt}
    \item Exp2: Both fullband DNN and wideband DNN$_{16}$ employ three sequentially stacked layers of LSTM\cite{lstm}, with the hidden units all set to 256.
      \vspace{-3pt}
\end{itemize}

PESQ-WB, STOI, SiSNR and SDR are employed as evaluation metrics, where PESQ-WB and STOI are only used for wideband evaluation. For wideband and highband evaluation, the fullband output signal of SE systems is first converted to the wideband and highband signal by a 0-8kHz low-pass filter and a 8-24kHz high-pass filter respectively.

  \vspace{-9pt}
\subsection{Backward compatibility for wideband enhancement}
\label{ssec:exp1-16}

\begin{table}
  \vspace{-6pt}
  \caption{0-16k results for Exp1 }
  \label{tab:encdec_pesq}
  \centering
  \small
  \setlength{\tabcolsep}{1mm}{
  \begin{tabular}{l | c c c c}
    \toprule
    \textbf{}     & \textbf{PESQ-WB} & \textbf{STOI} & \textbf{SiSNR} & \textbf{SDR} \\
    \midrule
     Noisy             & $2.00$ & $0.96$ & $8.32$ & $8.41$~~~          
     \\
     Mel$_{48}$         & $2.20$ & $0.95$ & $11.79$ & $13.45$~~~             \\
     Mel$_{64}$         & $2.55$ & $0.96$ & $14.41$ & $16.26$~~~             \\
     Mel$_{80}$         & $2.65$ & $0.97$ & $16.28$ & $18.27$~~~             \\
     FFT$_{768}$         & $2.39$ & $0.95$ & $14.23$ & $16.71$~~~ 
              \\
     TS-FFT$_{768}^{e16k}$   & $2.70$ & $0.97$ & $16.63$ & $19.44$~~~
              \\
    \bottomrule
  \end{tabular}}
  
\end{table}

\begin{table}
  \vspace{-6pt}
  \caption{16-48k \& 0-48k results for Exp1}
  \label{tab:encdec_fb}
  \centering
  \small
  \setlength{\tabcolsep}{1mm}{
  \begin{tabular}{l | c c | c c}
    \toprule
     \textbf{}  & \multicolumn{2}{c}{\textbf{highband(16-48k)}} & \multicolumn{2}{c}{\textbf{fullband(0-48k)}} \\ 
    \textbf{}     & \textbf{SiSNR} & \textbf{SDR} & \textbf{SiSNR} & \textbf{SDR}\\
    \midrule
     Noisy         & $-8.54$ & $-3.40$ & $8.27$ & $8.30$~~~         
           \\
     Mel$_{48}$         & $-1.97$ & $0.73$ & $11.75$ & $12.98$~~~             \\
     Mel$_{64}$         & $-0.91$ & $1.23$ & $14.33$ & $15.62$~~~             \\
     Mel$_{80}$         & $-0.84$ & $1.51$ & $16.16$ & $17.68$~~~              \\
     FFT$_{768}$         & $0.37$ & $2.73$ & $14.17$ & $16.14$~~~             \\
     TS-FFT$_{768}$   & $-9.66$ & $-1.27$ & $16.39$ & $18.51$~~~           \\
     TS-FFT$_{768}^{n16k}$   & $0.11$ & $2.19$ & $16.50$ & $18.57$~~~           \\     
     TS-FFT$_{768}^{e16k}$   & $2.14$ & $3.55$ & $16.51$ & $18.58$~~~           \\
    \bottomrule
  \end{tabular}}
    \vspace{-9pt}
\end{table}

The wideband (0-16k) SE performance is firstly evaluated as it provides the major contribution to the speech quality. As shown in Table \ref{tab:encdec_pesq}, for the wideband region, the proposed TS-FFT$_{768}^{e16k}$ condition (TS-FFT$_{768}$ and TS-FFT$_{768}^{n16k}$ has the same result in the wideband region) achieves the highest PESQ-WB, STOI, SiSNR and SDR scores. Compared to condition Mel$_{48}$, Mel$_{64}$, Mel$_{80}$, this result is reasonable as the frequency resolution of condition TS-FFT$_{768}^{e16k}$ is higher. Compared to condition FFT$_{768}$, TS-FFT$_{768}^{e16k}$ is more effective since a standalone model is used for wideband SE. When training the fullband system FFT$_{768}$ using the original STFT features, the feature dimension for the 8-24kHz frequency region is twice of the feature dimension for wideband region, resulting in sub-optimal performance in wideband region which is more perceptual important than the 8-24kHz frequency region. The results also justified the importance of using the psycho-acoustically motivated features in a one-step system. It should also be noted that the objective score of condition Mel$_{80}$ is only slightly lower than condition TS-FFT$_{768}^{e16k}$. This demonstrates the effectiveness of the common choice of the mel-scaled feature as in \cite{ping2018deep} and \cite{Chen2020HiFiSingerTH}.

  \vspace{-9pt}
\subsection{Superior performance for fullband enhancement}
\label{ssec:exp1-16-48}

The highband (16-48k) as well as the fullband (0-48k) speech quality is also evaluated. As shown in Table \ref{tab:encdec_fb}, TS-FFT$_{768}^{e16k}$ outperformed TS-FFT$_{768}$ and TS-FFT$_{768}^{n16k}$ in highband indicating the importance of the $\hat{S}_{16}$ to the DNN$_{16-48}$ network. The proposed two-step system TS-FFT$_{768}^{e16k}$ also outperformed other one-step systems for highband and fullband enhancement. Among these one-step systems, Condition FFT$_{768}$ achieved the highest score due to its biased frequency resolution toward the 8-24kHz region as discussed earlier.


\begin{table}
  \vspace{-6pt}
  \caption{0-16k results for Exp2}
  \label{tab:lstm_pesq}
  \centering
  \small
  \setlength{\tabcolsep}{1mm}{
  \begin{tabular}{l | c c c c}
    \toprule
    \textbf{}     & \textbf{PESQ-WB} & \textbf{STOI} & \textbf{SiSNR} & \textbf{SDR} \\
    \midrule
    Noisy             & $2.00$ & $0.96$ & $8.32$ & $8.41$~~~          
            \\
     Mel$_{48}$         & $2.11$ & $0.95$ & $11.69$ & $13.20$~~~                     \\
     Mel$_{64}$         & $2.50$ & $0.96$ & $14.42$ & $16.14$~~~                     \\
     Mel$_{80}$         & $2.55$ & $0.97$ & $16.24$ & $18.05$~~~                     \\
     FFT$_{768}$         & $2.40$ & $ 0.96$ & $16.20$ & $18.30$~~~ 
           \\
     TS-FFT$_{768}^{e16k}$   & $2.65$ & $0.97$ & $17.83$ & $20.07$~~~
           \\
    \bottomrule
  \end{tabular}}
  
\end{table}

\begin{table}
  \vspace{-6pt}
  \caption{16-48k \& 0-48k results for Exp2}
  \label{tab:lstm_fb}
  \centering
  \small
  \setlength{\tabcolsep}{1mm}{
  \begin{tabular}{l | c c | c c}
    \toprule
    \textbf{}  & \multicolumn{2}{c}{\textbf{highband(16-48k)}} & \multicolumn{2}{c}{\textbf{fullband(0-48k)}} \\ 
    \textbf{}     & \textbf{SiSNR} & \textbf{SDR} & \textbf{SiSNR} & \textbf{SDR}\\
    \midrule
     Noisy         & $-8.54$ & $-3.40$ & $8.27$ & $8.30$~~~         
     \\
     Mel$_{48}$         & $-0.63$ & $1.19$ & $11.65$ & $12.76$~~~            
     \\
     Mel$_{64}$         & $-0.58$ & $1.46$ & $14.35$ & $15.54$~~~            
     \\
     Mel$_{80}$         & $-0.22$ & $1.73$ & $16.14$ & $17.53$~~~             
     \\
     FFT$_{768}$         & $0.54$ & $2.89$ & $16.13$ & $17.73$~~~            
     \\
     TS-FFT$_{768}$   & $-9.66$ & $-1.27$ & $17.17$ & $18.89$~~~           \\
     TS-FFT$_{768}^{n16k}$   & $0.11$ & $2.19$ & $17.32$ & $18.95$~~~           \\
     TS-FFT$_{768}^{e16k}$   & $2.06$ & $3.51$ & $17.33$ & $18.96$~~~           \\
    \bottomrule
  \end{tabular}}
  \vspace{0pt}
\end{table}

  \vspace{-9pt}
\subsection{Generalization on different network structures}
\label{ssec:exp2}

\begin{figure}[!htb]
  \centering
  \includegraphics[width=0.85\linewidth]{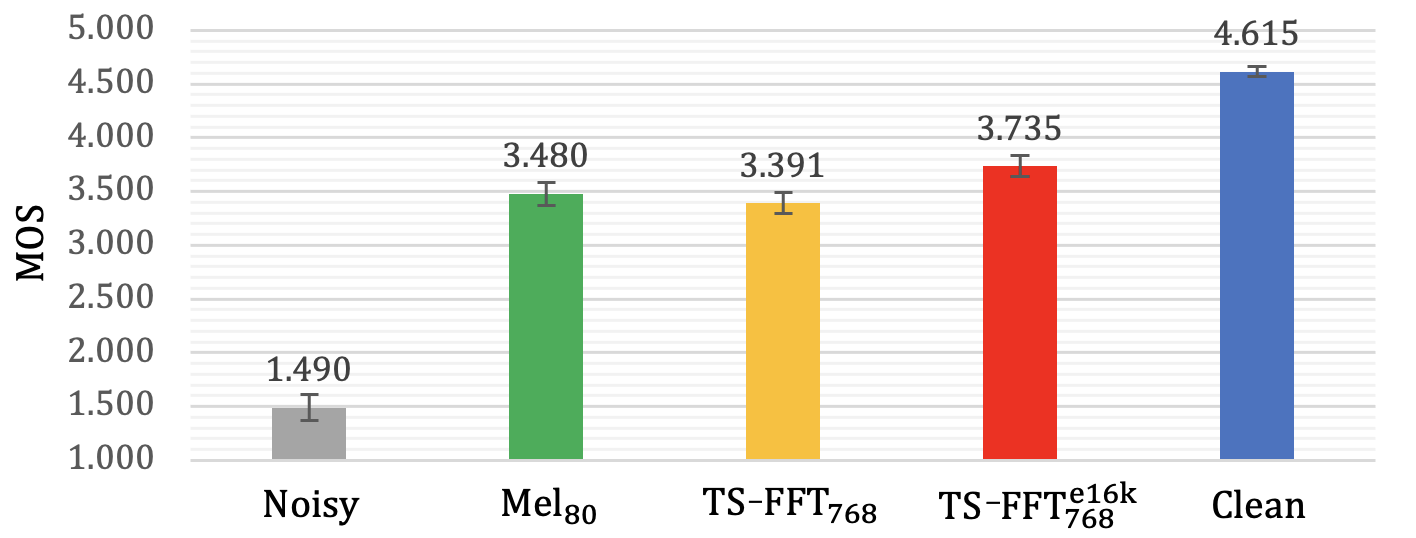}\\
  \vspace{-10pt}
  \caption{Subjective Mean Opinion Score Test}
  \label{fig:MOS}
  \vspace{-12pt}
\end{figure}
To further evaluate the generalization ability of the proposed two-step fullband SE system, the encoder-decoder structure in Section \ref{ssec:exp1-16} and \ref{ssec:exp1-16-48} is replaced by three stacked LSTM layers. The identical evaluation as in Table \ref{tab:encdec_pesq} and \ref{tab:encdec_fb} is performed. The results are listed in  Table \ref{tab:lstm_pesq} and \ref{tab:lstm_fb}. 
As shown, similar conclusions as in Section \ref{ssec:exp1-16} and \ref{ssec:exp1-16-48}  can be reached using the stacked LSTM structure.

\vspace{-6pt}
\subsection{Subjective Mean Opinion Score Test}
\label{ssec:exp3}
\vspace{-6pt}
A subjective MOS test is also conducted with 10 randomly selected noisy speech files processed by 3 of the representative conditions in Table 4. The noisy and the clean speech conditions are also considered resulting 5$\times$10=50 48kHz audio samples. 15 listeners participated the MOS test. The results with 95\% confidence interval are shown in the Figure \ref{fig:MOS}. As shown, the proposed system (Condition TS-FFT$_{768}^{e16k}$) achieved statistically higher subjective MOS score than the one-step system (Mel$_{80}$) and the two-step system without using the aid of $\hat{S}_{16}$ (TS-FFT$_{768}$), demonstrating the effectiveness of the proposed system in Figure \ref{fig:NetworkArchitecture} (b).

\vspace{-3pt}
\section{Conclusions}
\label{sec:conclusions}
\vspace{-6pt}
A two-step backward compatible fullband speech enhancement system is proposed. Compared with the existing one-step structure, the proposed system can achieve comparable performance for the wideband speech quality while significantly outperformed the existing structure for the highband speech quality and ensured backward compatibility with the existing wideband SE approaches. 

\bibliographystyle{IEEEbib}
\bibliography{strings,refs}
\end{document}